\documentstyle[12pt]{article}
\newcommand{\be}{\begin{equation}}
\newcommand{\ee}{\end{equation}}
\newcommand{\bea}{\begin{eqnarray}}
\newcommand{\eea}{\end{eqnarray}}
\newcommand{\nn}{\nonumber}

\newcommand{\bfpi}{\mbox{\boldmath $\pi$}}

\begin{document}

\def\theequation{\arabic{section}.\arabic{equation}}
\def\thesection{\Roman{section}}
\def\thesubsection{\Alph{subsection}}

\setcounter{section}0
\setcounter{subsection}0
\setcounter{equation}0

\hfill{JINR E2 - 95-203}

\hfill{ZU - TH-4/95}

\vspace*{1cm}

\begin{center}
{\large \bf                 ON    ADMISSIBLE     GAUGES
\\
                                       FOR
\\\vspace{0.25cm}
                           CONSTRAINED         SYSTEMS}
\end{center}

\bigskip
\bigskip

\centerline{\sf S. A. Gogilidze, \footnotemark[1] ~
A. M. Khvedelidze, \footnotemark[2]~  V. N. Pervushin}

\bigskip

\centerline{
\it Joint Institute for Nuclear Research , 141980, Dubna, Russia}
\bigskip
\bigskip

\sf
\thispagestyle{empty}

\footnotetext[1]{
\normalsize
Permanent address: Tbilisi State University,
380086, Tbilisi, Georgia.}

\footnotetext[2]{
\normalsize
Permanent address: Tbilisi Mathematical Institute,
380093, Tbilisi, Georgia.\\
\hspace*{0.63cm} Electronic address: khved@theor.jinrc.dubna.su}

\bigskip
\bigskip
\bigskip
\bigskip
\bigskip
\begin{abstract}

The {\it {gauge - fixing} } and  {\it gaugeless } methods
for  reducing the  phase space  in the
generalized  Hamiltonian dynamics are compared
with the aim to  define the  class of admissible gauges .

In the gaugeless approach,  the reduced phase space of a
Hamiltonian system with the first class constraints is constructed
locally,  without  any gauge  fixing, using
the following procedure:  abelianization of
constraints  with  the subsequent   canonical transformation
so that some of the new momenta are equal to the new abelian constraints.
As a result  the corresponding conjugate coordinates are ignorable (
nonphysical )
one while  the  remaining  canonical pairs corresponds to the true dynamical
variables.
This representation for the phase space  prompts us the definition of
subclass
of admissible gauges --- canonical gauges  as  functions
depending only on the ignorable coordinates.
A practical method to recognize the canonical gauge is proposed .
\end{abstract}

\newpage

\author{\sf S. A. Gogilidze,
 ~A. M. Khvedelidze,
{}~  V. N. Pervushin\\
}


\title{\bf  ON ADMISSIBLE  GAUGES  FOR  CONSTRAINED  SYSTEMS  }

\date{}
\maketitle
\sf
\thispagestyle{empty}



\bigskip
\bigskip

\tableofcontents

\newpage
\vspace*{2cm}

\setcounter{page}{1}

\section{\sf Introduction}

It is the purpose of this note to discuss  the problem
of constructing  of `` true dynamical degrees of freedom''
in the degenerate theories with first class
constraints  with the aim
to obtain a constructive definition of  admissible gauges.

A general  method for describing  degenerate theories starts by
introducing the gauge fixing condition ( gauge  fixation )
for elimination of nonphysical degrees of freedom ~\cite{Dirac59} -
\cite{Sunder}.
Afterwards, there are two significantly  different ways  for  reduction
in number of degrees of freedom : explicit and implicit .
The explicit method is  straightforward -- one must deal in an explicit way
only with the physical variables  while nonphysical ones are  completely
excluded from the consideration  by gauge fixation.
For the Abelian gauge theories this method works,  one can always
find the gauge invariant
variables  and identify them with physical ones while the rest variables
can be dropped out  via the  gauge conditions.
A problem arises, for example, in the
Yang - Mills theory and gravity, where  the true dynamical degrees of
freedom  are hidden  due to the non - Abelian  character of theories.
The problems concerning the determination of the
physical degrees of freedom in the Yang - Mills theory
have been discussed by many authors
(see e.g. \cite{GoldJack} - \cite{Khved}).
A lot of attempts have been undertaken to
realize the explicit separation of  the nonphysical sector from the
physical one.
However, there still remain open questions and as a rule  in the  practical
calculations  we  deal with the implicit form of gauge fixation.
In this case, the general method ~\cite{Dirac59},
providing the restriction on the canonical variables  due to  constraints
(including  gauge fixing)  consist in the
determination of
the modified symplectic structure of the  phase space
with  the help of Dirac 's bracket  ~\cite{DiracL} .
In this method,  one retains all dynamical variables
and merely  changes  their Poisson brackets,  which corresponds
to the effective  reduction in the
number of  degrees of freedom.
As a result, one could not in general indicate  the coordinates  of
the reduced system corresponding to the  `` true dynamical  degrees  of
freedom ''.
However, to attain the correct results, one must be sure that the gauge
fixing condition allows one
to eliminate nonphysical degrees of freedom  and  fix
the physical one in a unique way ( up to canonical transformation ) without
any restriction on them.
One can ask the question : are there some  requirements to the  gauge
fixing conditions that  quarantee  such a
correct description .

It is important to note that there are  two  sides of this question :
global and  local \cite{Sunder} .
The well known manifestation of the problem of determining of
globally admissible gauges
is the  so called  Gribov ambiguity for the Yang - Mills theory.
\cite{Gribov}.
Singer's no - go theorem for gauge fixing \cite{Singer}
rises  questions about  the generalization of the usual procedure of
reducing  degrees of freedom based on local manipulations .
However,  as we want to emphasize, at least  for a local procedure it is
necessary to  clarify  the reduction scheme.
Just this is the goal of the  present paper.

According to  Dirac's prescription  for
generalized Hamiltonian systems, the  reduction in the number of
degrees of freedom
consist in the elimination of  first class constraints
\[
   \varphi_\alpha (p,q) \,=\,0
\]
by  introducing  some new ``gauge constraints'' into the theory
\[
\chi_\alpha (p,q) = 0 \,
\]
and by  replacing  of the Poisson bracket by
the Dirac one ~\cite{Dirac59}.
The gauge functions \(\chi\)  are   arbitrary functions
of coordinates and momenta. There is only one requirement on the gauge
fixing  condition --- nonvanishing of the Faddeev - Popov determinant
on the constraint ( including  gauge fixing ) shell
 \be \label{eq:det}
\framebox[95mm]{ \raisebox{0ex}[4ex][3ex]{
$ \det \Vert {\{\chi_\alpha(p,q), \varphi_\beta (p,q)\}} \Vert \:
\Bigl\vert_{ \varphi = 0,\: \chi = 0 }\: \not=  \,0 $}}
\ee
However, it is  known that (\ref{eq:det}) is only a necessary condition for
the gauge constraints \cite{Sunder}. There are examples of gauge constraints
fulfilling
(\ref{eq:det}), but  as a result of reduction we get some restriction on
physical sector --- some type of overconstraining.
For  explanation, let us consider the simple case of
QED with constraints
\[
\varphi_1 = \pi_0  \,, \quad  \varphi_2 =\partial_i \pi_i
\]
If one chooses the following gauge
\[
\chi_1 = A_0 = 0 \,, \quad  \chi_2 = A_3 = 0,
\]
then the Faddeev - Popov matrix
\( \{\chi_\alpha, \varphi_\beta\}  \) has a non - singular determinant
on appropriate function space.
But it is easy to state
that this gauge leads to overconstraining  of the system .
Indeed, according to the essence  of gauge fixing ---  to get rid
of some degrees of freedom, the gauge - fixing  condition   allows  one
to determine in a unique manner the gauge transformation function from
this gauge  fixing conditions \cite{FadSlav} :
\bea
A_0^\lambda & \equiv & A_0 + \partial_0 \lambda = 0  \,\nonumber\\
A_3 ^\lambda & \equiv & A_3 + \partial_3  \lambda = 0 .\nonumber
\eea
It is obvious that there is a unique  solution  to these equations
with respect to \(\lambda\) if the integrability condition is satisfied
\[
\partial_3 A_0 - \partial_0  A_3  \equiv  0
\]
Thus, we obtain  the restriction on the physical variable, the  third
component of the electric field
\[
\pi_3 = 0,
\]
and get the overconstraining of QED. ( see e.g
\cite{Sugano}, \cite{McMullan} ).

So,  to be sure that we are free from some incompatibility, it would be
ideal if one could pick out directly the degrees of freedom
(whitch are unconstrained) that have to be dropped out  from a set of
canonical pairs and then one would work in the reduced phase space .
In other words, to get some information on restriction of gauge conditions,
it is necessary to  deal with some scheme that allows
us to determine the reduced dynamics in a gaugeless manner  and then
to compare it with gauge fixing method .
Fortunately, there is an elegant method  of reducing the number of
degrees of freedom  known for systems of equations in involution
\cite{Levi-Civita}, \cite{Whittaker}.
Levi- Civita has proposed  the way of
using the invariant relations (constraints in modern notations )  to
reduce the order of the canonical system by passing to new canonical
variables.
As a result of the application of this scheme
\cite{Shanmugadhasan} - \cite{Newman},
the  new canonical variables in the reduced system describe the
allowed dynamics in terms of physical variables.
It should be noted that for a direct application to the non - Abelian theory
and gravity there is a serious  obstacle.  In this case, before  carring
 out the canonical transformation to new variables, the constraints
must be replaced
by the equivalent set of constraints that  form a canonical functional
group. There is a general proof of a possibility  of such a replacement
\cite{Kulk} - \cite{HenTeit}, but
the problem is to  determine this new set in a constructive fashion.
Nevertheless,  this gaugeless scheme allows one to obtain
some restriction on gauge fixing conditions .

In the present paper, based on the gaugeless scheme of reduction of
the phase space we suggest
in a constructive manner a certain subclass of admissible gauges
(canonical gauges)  for gauge theories with a first class constraints
which can be exploited in the gauge fixing method.
One can note  a  simple condition for  gauge fixing functions
which can  serves a criterion  for belonging to the class of canonical
gauges ---  the requirement of
vanishing  the Dirac bracket of matrix \( \Delta_{\alpha \beta } =
\{\chi_\alpha, \varphi_\beta \} \)  with the canonical Hamiltonian on
 the constraint (including  gauge fixing) shell
\be \label{eq:sufcon}
\framebox[90mm]{ \raisebox{0ex}[4ex][3ex]{
$ \{ \Delta_{\alpha \beta} (p,q) , H_C(p,q) \}{}_{D}\;
\Bigl\vert_{ \varphi = 0, \:\chi = 0 }\,
= 0 $ } }
\ee

This  article is  organized as  follows.
In the first part of this paper  we shall briefly describe
 Dirac's  gauge fixing method and the gaugeless one .
Section II  is  devoted  to  the definition of admissible gauges
based on the canonical equivalence
between  two methods. In the last  section,
the  general consideration of the admissible gauges
is  exemplified by
Christ and Lee model ~\cite{Christ}.

\section{\sf Phase space  of the Hamiltonian system with constraints }

\bigskip

For the sake of simplicity, as usual we will discuss the main ideas
using mechanical system,  i.e. system  with a  finite number of degrees of
freedom, with having in mind that the transition to a field theory
involves additional features connected with boundary effects.

\subsection{\sf Definition of the reduced phase space}

Suppose that in the  system with finite number of degrees of freedom
we have  the following first class constraints
\bea                \label{eq:constr}
   \varphi_\alpha (p,q) \,& = & \,0 , \nn\\
 \{ \varphi_\alpha (p,q),  \varphi_\beta  (p,q)\}\,&  = &\,
f_{ \alpha\beta \gamma}  (p,q) \varphi_\gamma  (p,q).
\eea
This means that the dynamics of our system is constrained on a certain
submanifold of the total phase space
which is defined by the  constraints \( \varphi_\alpha (p,q) \)
in (\ref{eq:constr}).
Further, we will symbolize by notation \({\Gamma}_{c} \)
this \( 2n - m \) --- dimensional submanifold
of the total phase space \( {\Gamma} (\, dim \Vert \Gamma \Vert =  2n \,) \)
\(
{\Gamma}_{c}   \subset {\Gamma} .
\)
For definition of the reduced or physical phase space we need  the notion
of physical variable .  According to  Dirac :
{\it  A dynamical variable \(  F \)  is of physical importance only if its
Poisson bracket with any constraints gives another
constraint }\cite{Dirac49}
\be                \label{eq:physvar}
     \{ F (p,q ), \varphi_\alpha (p,q), \}  =
d_{ \alpha \gamma} (p,q) \varphi_\gamma  (p,q).
\ee
Such a dynamical variable  is called a {\it physical variable }.
According to this definition,  in the process of evolution a
physical variable  does not abandon some subspace of \( {\Gamma}_{c}\).
Indeed ~\cite{Faddeev69}, \cite{KonPop},  if one consider, (\ref{eq:physvar})
as a set of \( m \) first order linear differential equations for
\( F \), than due to the integrability condition (\ref{eq:constr})
this function  can be  completely determined by its values in the
\( 2(n-m) - m \) submanifold of its initial conditions.
Thus, observables are functions on the socalled reduced phase
submanifold \( {\Gamma}^{\ast} \)
\(
{\Gamma}^{\ast} \subset  {\Gamma}_{c}   \subset {\Gamma}
\)
spanned by some physical coordinates
\(Q^{\ast}_i,\: P^{\ast}_i \:(i = 1, ... , 2(n-m)) \).
Below we will discuss  alternative schemes of construction of
the reduced phase space: gauge  - fixing and gaugeless methods.

\subsection{\sf Reduced phase space in  the gauge fixing method}
\bigskip

\subsubsection{\sf Dirac's scheme without constraint resolution}
\bigskip

Let us briefly describe the general principles of the
introduction of gauge fixing constraints  on  canonical variables
in a Hamiltonian theory.
This  general procedure to deal with physical variables
was proposed by
Dirac for the application to the Hamiltonian theory of gravitation
\cite{Dirac59}.

The generalized  Hamiltonian dynamics is described by the extended
Hamiltonian
\be  \label{eq:exham}
H_E (p,q) = H_C(p,q) + u_\alpha (t) \varphi_\alpha (p, q) \,
\ee
where \( H_C (p,q)\) is the canonical Hamiltonian  and \( u_\alpha \) are
the Lagrange multipliers .
According to Dirac's  gauge fixing prescription, one can introduce the new
``gauge '' constraints
\be
\chi_\alpha (p,q) = 0                  \label{eq:gauge2}
\ee
with  the requirement
\be
\det \Vert \{\chi_\alpha (p,q), \varphi_\beta (p,q)\} \Vert \, \not= \, 0.
\ee
The maintenance of auxiliary conditions (\ref{eq:gauge2}) in time gives
the set of equations
\be
 \dot{\chi}_\alpha = \{\chi_\alpha, H_C \} + \sum_{\beta} \{\chi_\alpha,
\varphi_\beta  \} u_\beta = 0
\ee
which allows to determine the unknown Lagrange multipliers.
Formally, the solution can be written as
\be
u_\alpha =  - \sum_{\beta} \Delta^{-1}_{\alpha \beta}\{H_C, \chi_\beta \} \,
\ee
where  \(\Delta^{-1} \) is the inverse  matrix of
\[
\Delta_{\alpha \beta} = \{\chi_\alpha , \varphi_\beta\} ,  \quad
\Delta_{\alpha \beta}  \Delta^{-1}_{\beta \gamma} = \delta _{\alpha \gamma} \]

The main idea of introduction the new constraints (\ref{eq:gauge2})
into the theory was to eliminate from consideration
the complicated constraints  (\ref{eq:constr}),
i.e.
to consider them  as strong equations.
This result can be achieved if we pass from  the Poisson brackets
to  Dirac's ones
\be  \label{eq:DB}
\{F , G \}{}_{D} \equiv  \{F, G \} -
\{F ,\xi _s\}C^{-1}_{s s`} \{ \xi_{s`}, G \},
\ee
\[
\xi_s \equiv \left( \varphi_1 , \dots , \varphi_m ,
\chi_1, \dots , \chi_m \right), \quad
C_{\alpha \beta} \equiv \{\xi_\alpha , \xi_\beta\},   \quad
C_{\alpha \beta}  C^{-1}_{\beta \gamma} = \delta _{\alpha \gamma} \]

 From (\ref{eq:DB}) one can observe that  all constraints including the
gauge one have zero Dirac's brackets  with everything  and thus we
can consider them as strong equations.
As it has been  mentioned in the introduction, although the choice of
gauge constraints
allows one to take into account in an explicit form the constraint nature of
canonical variables via Dirac's bracket  but this gauge fixing does not
provide
an explicit  representation for the physical phase space.
We will deal with the explicit representation for the reduced phase space if
 one can find the conjugate  coordinates \(Q^{\ast}_i, P^{\ast}_i
(i = 1, \dots , n-m \), so that all constraints would vanish identically as
functions of these  variables \( \varphi_\alpha (p, q) \equiv
 \overline{\varphi}_\alpha  (Q^{\ast}_i P^{\ast}_i ) \equiv 0 \)
 \cite{Sunder}.
In this case, for any function  \( F(p,q) \) given on the
reduced phase space
\[
 F(p,q)
\Bigl\vert_{ \varphi = 0 \chi = 0} \, = \overline{F}(P^{\ast} Q^{\ast})
\]
the Dirac bracket looks like the Poisson bracket for a usual
unconstrained system
\be  \label{eq:PDB}
\{F,G \}_{D}
\Bigl\vert_{ \varphi = 0 \chi = 0}
=  \sum_{i=1}^{n-m}\left\{
\frac{\partial \overline{F}}{\partial Q^{\ast}_i}
\frac{\partial \overline{G}}{ P^{\ast}_i} -
\frac{\partial \overline{F}}{\partial P^{\ast}_i}
\frac{\partial \overline{G}}{ Q^{\ast}_i}\right\}
\ee
However, it is not  easy to find these coordinates and
in general one retains all dynamical variables.
The change  of their Poisson brackets  reflects the reduction in
number of the degrees of freedom
\[
\sum_{i =1}^{n}
\{q_i, p_i, \}_{P.B.} = n, \quad
\sum_{i =1}^{n}
\{q_i, p_i, \}_{D}
= {n-m}
\]
Thus, the question of `` true dynamical  degrees ''
is again open.

\subsubsection{\sf Faddeev's  scheme with constraint resolution}
\bigskip

In the well - known paper by L.D. Faddeev \cite{Faddeev69}
the scheme of explicit reduction of phase space
with the goal to
extend the method of path integral quantization to a gauge theory
was developed.
Here we will stress only the main points of this scheme.
As in Dirac's methods, we  introduce the constraints
\[
\chi_\alpha (p,q) = 0 \,
\]
in such a way that the requirement (\ref{eq:det}) is fulfilled
with an additional property
\be                \label{eq:abg}
     \{ \chi_\alpha (p,q), \chi_\beta(p,q) \}  =  0.
\ee
Now, in accordance with this property there is a  canonical
transformation to new coordinates
\bea
q_i & \mapsto & Q_i = Q_i \left ( q_i , p_i \right )\nn\\
p_i & \mapsto & P_i = P_i \left ( q_i , p_i \right )
\eea
such  that  \( m \) of the new  \( P  \) 's
are
\be
{P}_\alpha = \chi _\alpha \left ( q_i , p_i \right )
\ee
The corresponding  conjugate variables \( Q_\alpha\)
can be  expressed
with the  help of the resolution of constraints
(\ref{eq:constr})
 \[
Q_\alpha = Q_\alpha \left ( {Q}^{\ast}, {P}^{\ast}  \right)
\]
via the  \( n - m \) cannonical pairs
(\(
{Q}^{\ast}_1, {P}^{\ast}_1,\dots ,
{Q}^{\ast}_{n-m}, {P}^{\ast}_{n-m} \)).
This is possible due to the (\ref{eq:det}).
This remaining variables
 ( \(
{Q}^{\ast}_1, {P}^{\ast}_1
 \dots ,
{Q}^{\ast}_{n-m}, {P}^{\ast}_{n-m} \) )
spanned the \(2(n-m) \) - dimensional surface \(\Sigma\)
determined  by the equations
\bea
P_\alpha & = & 0 \nn\\
Q_\alpha & = & Q_\alpha \left ( {Q}^{\ast}, {P}^{\ast}  \right)
\eea
After this the main point is to prove that the surface
\(\Sigma\) coincides with the true reduced phase space
\({\Gamma}^{\ast} \)  independently of the choice of
gauge fixing condition.
It is very attractive to determine
the reduced phase \({\Gamma}^{\ast} \) without any gauge fixing and then
to compare  it with the reduce phase space obtained by  Faddeev's
gauge fixing method.
In the next section we will describe two schemes of reduction
of phase space without exploiting gauge fixing functions, solely
in internal terms of the theory.

\subsection{\sf Construction of  the reduced phase space
without gauge fixing
 via  the ~~ ``generalized
canonical transformation''}

\subsubsection{\sf Abelian constraints }
\begin{center}
{\it i. Levi - Civita's method of reduction of
systems in involution}
\end{center}

For explanation of main ideas of construction
of the  reduced subspace  \({\Gamma}^{\ast} \)
without using gauge fixing condition i.e . in the gaugeless manner
let us first consider the special case when there are only
Abelian constraints in the theory
\be                \label{eq:ab1constr}
     \{ \varphi_\alpha (p,q),  \varphi_\beta  (p,q) \}  = 0.
\ee
In this case a difficulty does not arise  because there  is a
general method of reducing in order of differential equation in the
canonical
if  some invariant  relations in involution  are  known
{}~\cite{Levi-Civita},
\cite{Whittaker}, \cite{Shanmugadhasan}.
According to Levi - Civita's method, one can  perform the canonical
transformation in the phase space to the new coordinates
\footnote{
{
Note , that  in the case of the theory with reparametrization invariance  we
have to  exploiti a more general
transformation with an explicit time dependence.
But in this article we restrict ourselves only to the
case of gauge invariant theories. }}
\bea \label{eq:cantr}
q_i & \mapsto & Q_i = Q_i \left ( q_i , p_i \right )\nn\\
p_i & \mapsto & P_i = P_i \left ( q_i , p_i \right )
\eea
such  that  \( m \) of the new  \( P  \) 's
 ( \(\overline{P}_1, \dots ,\overline{P}_m \) )
become equal to the constraints
(\ref{eq:ab1constr})
\be
\overline{P}_\alpha = \varphi _\alpha \left ( q_i , p_i \right )
\ee
while the remaining  \( n-m \) pairs of the new canonical coordinates
(\(
{Q}^{\ast}_1, {P}^{\ast}_1
 \dots , \\
{Q}^{\ast}_{n-m}, {P}^{\ast}_{n-m} \) )
will be  gauge invariant physical variables.

In terms of new canonical pairs \(P, Q \) it is very useful to  establish the
general structure
of the canonical Hamiltonian.
The maintenance of complete system of irreducible constraints
(\ref{eq:ab1constr}) in time means that
\be \label{eq:maconstr}
\{\varphi_\alpha (p,q), H_C (p,q) \} = g_{\alpha\beta} (p,q)
\varphi_\beta (p,q)
\ee
Eq. (\ref{eq:maconstr}) in the new coordinates \(P, Q\) becomes
\be \label{eq:smaconstr}
\frac{\partial{\overline{ H}_C(P,Q)}}{\partial{\overline{Q}_\alpha}}
=\overline{g}_{\alpha\beta}(P,Q)\overline{P}_\beta
\ee
\[ \overline{ H}_C(P,Q)  =  H_C(p(P,Q), q(P,Q))  \]
 From this equation it follows that  the canonical  Hamiltonian
\( H_C (p,q) \), rewritten in the new coordinates  \( P, Q \)
has the  following form
\be  \label{eq:hamrep}
\overline{H}_C (P,Q) =\overline{H}_0({Q}^{\ast}, {P}^{\ast}, \overline{P})
 + \overline{\Psi}_\alpha (Q,P) \overline{P}_\alpha
\ee
with some  function \( H_0( {Q}^{\ast}, {P}^{\ast},  \overline{P})\)
which does not depend  on the ignorable coordinate
\( \overline{Q}\) :
\be
\{\overline{P}_\alpha, \overline{ H}_0(P,Q) \} = 0 \quad \Longrightarrow
 \quad   \{\varphi_\alpha(p,q), { H}_0(p,q) \} = 0
\ee
and thus represents the gauge invariant part of canonical Hamiltonian.
The functions \(
\overline{\Psi}_\alpha (Q,P)\)
are determined through the functions \(\overline{g}_{\alpha\beta}(P,Q)\)
according to the  equation
\be \label{eq:spsi}
\frac{\partial{\overline{\Psi}_\gamma(P,Q)}}{\partial{\overline{Q}_\alpha}}
=\overline{g}_{\alpha \gamma}(P,Q)
\ee
This property in the initial  coordinates \( p,q \)
means that , the canonical Hamiltonian looks as follows:
\be
H_C (p,q) = H_0(q,p)
 + \Psi_\alpha (p,q) \varphi_\alpha (p,q)
\ee
whith the  gauge invariant  function
\( H_0 (p,q)\)
\be
\{ H_0 (p,q ), \varphi_\alpha (p,q) \} = 0  \label{eq:invham}
\ee
and  functions \(
\Psi_\alpha (p,q) \) connected with   \( \overline{\Psi}_\alpha (Q,P)\)
as
\be
\Psi_\alpha (p,q) = \overline{\Psi}_\alpha (P(p,q), Q(p,q))
\ee
Eq.(\ref{eq:spsi}) rewritten in canonically invariant form looks like
\be
\{
\varphi_\alpha(p,q), \Psi_\gamma(p,q)\} = {g}_{\alpha\gamma}(p,q)
\ee
One would like to note that
the simple definition  of invariant part of canonical Hamiltonian
in terms  of special coordinates
\be
H_{0} (P, Q) =  \overline{H}_C (P,Q) -
\frac{\partial\overline{H}_C}{\partial\overline{P}_\alpha}
\overline{P}_\alpha
\ee
in the old coordinates can be written
only  through the variational derivative
\be
H_{0} (p, q) = \left[ H_C(p,q) -
\frac{\delta H_C}{\delta \varphi_\alpha}\varphi_\alpha
\right]
\label{eq:physoham}
\ee
According to the Dirac, the time evolution of  a singular hamiltonian
system is  governed by the extended Hamiltonian (\ref{eq:exham})
\bea
 \dot{q}_i & = & \{ q_i, H_{E} \} \nn\\
 \dot{p}_i & = & \{ p_i, H_{E} \}
\eea
In the new special coordinates instead of these equations
we have the following factorazible form of
canonical equations for two type of variables
\( {P}^{\ast}_i, {Q}^{\ast}_i \;\; (i=1,\dots, n-m) \) and \(
\overline{P}_\alpha,
\overline{Q}_\alpha \;\; (\alpha = 1,\dots, m) \):
\cite{Shanmugadhasan}, \cite{Dominici}
\bea \label{eq:heq}
 \dot{Q}^{\ast}_i & = & \{{Q}^{\ast}_i, H_{Ph} \} \nn\\
 \dot{P}^{\ast}_i & = & \{{P}^{\ast}_i, H_{Ph} \} \nn\\
 \dot{\overline{P}}_\alpha & = & 0 \nn\\
 \dot{\overline{Q}}_\alpha & = & \overline{u}_\alpha(t)
\eea
with an arbitrary functions \( \overline{u}_\alpha(t) \).
In (\ref{eq:heq}) the physical Hamiltonian  is defined as
\be
H_{Ph} (P, Q) = H_0( {Q}^{\ast}, {P}^{\ast},  \overline{P})
 \Bigl\vert_{  \overline{P} = 0}
\label{eq:physham}
\ee
Thus,  \( \overline{Q}_\alpha \) are  ignorable coordinates with the
corresponding vanishing  momenta \(\overline{P}_\alpha\),  and the
canonical system  allows the separation of the phase space
coordinates  into the physical sector and the nonphysical one
\be \label{eq:1str}
2n \left \{ \left (
\begin{array}{c}
q_1\\
p _1\\
\vdots\\
q_n\\
p _n\\
\end{array}
\right )  \right.
\quad    \mapsto  \quad
\begin{array}[c]{r}
{ 2(n-m)\,\, \left\{ \left(
\begin{array}{c}
Q^\ast \\
P^\ast
\end{array}
\right ) \right. }\\
\vphantom{\vdots} \\
{ 2m\, \left\{ \left(
\begin{array}{c}
\overline{Q}\\
\overline{P}
\end{array}
\right ) \right. }
\end{array}
\quad
\begin{array}[@]{c}
{ \begin{array}{c}
 Physical \\
sector
\end{array}
} \\
\vphantom{\vdots} \\
{ \begin{array}{c}
Nonphysical\\
sector
\end{array}}
\end{array}
\ee
One would like to note that the choice of special canonical
coordinates
 \({P}^{\ast}, {Q}^{\ast}\) and \(\overline{Q}\) is not a unique.
It is a wide freedom to define them, for example one can pass to new
canonical variables
\bea \label{eq:free}
{\overline{P'}}_\alpha & = & \overline{P}_\alpha \nn\\
{\overline{Q'}}_\alpha & = & \overline{Q}_\alpha +
f_\alpha({Q}^\ast)\nn\\
{P'}_i^{\ast} & = & {P}_i^{\ast} + {\overline{P}}_\alpha
\frac{\partial f_\alpha({Q}^\ast)}{\partial {Q}_i^{\ast}} \nn\\
{Q'}_i^{\ast} & = & {Q}_i^{\ast},
\eea
but in any case the above redefinition
corresponds to the canonical transformation on  the physical
phase space spanned by the \(({Q}^{\ast}, {P}^{\ast}) \).
\bigskip
\bigskip
\subsubsection{\sf Non - Abelian constraints}

\bigskip

 If there are in the theory a non - Abelian constraints, the above -
described
procedure does not work and it is necessary to modify it.
Fortunately, there is the significant  observation that allows us to lead
this case  to the previous Abelian one.
It  can be note that in contrast with   unconstrained systems the  singular
theories  possess a  wider freedom in the choice of canonical
variables \cite{Bergman} .
The canonical group of transformation for a usual unconstrained theory
in this case is enlarged to the  group
of so - called `` generalized canonical transformations ''
According to the definition, the generalized canonical transformations
are  those preserving the form of all constraints
of the theory as well as the canonical form of the equations of motion
{}~\cite{Bergman}.
The main point of our idea is to use this freedom of formulation
and to pass from non - Abelian theory to an equivalent description
of singular system with Abelian constraints via the generalized canonical
transformation.
The usual canonical transformation of variables
could not  change the value of the Poisson brackets, but as we  will
demonstraed below the transformation to new Abelian constraints is not
canonical but  generalized canonical transformation.
In this section, we will  consider  two  schemes of
realization of this program based on the resolution of
constraint or  without it.

\bigskip
\begin{center}

{\it i. Construction of physical coordinates  via the constraint resolution
}

\end{center}
\bigskip

The direct way to pass to new constraints that are Abelian and
simultaneously are equivalent to
the  old one ( abelianization ) is  as follows~
\cite{Newman}, \cite{HenTeit}.
Under the assumption that \(\varphi_\alpha (p,q)  \)  are \( m \)
independent functions one can resolve the constraints
(\ref{eq:constr}) for \(m \) of \( p\)'s
\be
p_\alpha = F_\alpha (\underline{p}, q)
\ee
where \(\underline{p} \) denotes the remaining  \( p\)'s.
One can now define the new equivalent to \(\varphi_\alpha (p,q)  \)
constraints
\be \label{eq:modcon}
\Phi_ \alpha (p,q ) = p_\alpha  -  F_\alpha (\underline{p}, q)
\ee
Now, on the  one hand, by explicitlly computing one can convince  onself
that the Poisson brackets  \( \{\Phi_ \alpha (p, q)
 ,\Phi_ \beta (p,q ) \} \) of the new constraints are  independent of
\(p_\alpha\), but on the  another hand, they
are again the first class ones;
so  their Poisson brackets
with each other must vanish identically .
Thus after a transfomation  to  new constraints
 \( \Phi_ \alpha (p, q)  \)
we are ready to
realize the above mentioned canonical
 transformation (\ref{eq:cantr}) such  that  \( m \) of the new  \( P  \) 's
become
 equal to the modified constraints  \( \Phi_\alpha \)
(\ref{eq:modcon})
\be
\overline{P}_\alpha
 = \Phi_\alpha \left ( q_i , p_i \right )
\ee
with the corresponding conjugate ignorable coordinates
\( \overline{Q}_\alpha \) .

\bigskip

\begin{center}
{\it i.i. Abelianization of constraints via
Dirac's transformation }
\end{center}
\bigskip

There is another method of abelianization  without using
a non painless,  in general procedure of resolution of constraints
against some momenta.
In the previous article \cite{Gogil1}  it was shown
how due to the freedom in
the representation  of the constraint surface \( \Gamma_c \)
defined  by
\[
   \varphi_\alpha (p,q) \, =  \,0 ,
\]
with
\[
 \{ \varphi_\alpha (p,q),  \varphi_\beta  (p,q)\}\,  = \,
f_{ \alpha\beta \gamma}  (p,q) \varphi_\gamma  (p,q).
\]
one can always pass with the help of Dirac 's transformation
(which belongs to the class of generalized canonical transformations)
>from  these  first class non - Abelian  constraints
to the equivalent ones
\be                \label{eq:Diractran}
\Phi_\alpha (p,q) \, = {\cal D}_{\alpha \beta} (p, g) \varphi_\beta  (p,q)
\ee
so that new constraints are Abelian.
\bea                \label{eq:abconstr}
    \{ \Phi_\alpha (p,q),  \Phi_\beta  (p,q)\}\, &  = &\, 0.
\eea
As it has been mentioned above, the  existences of a such a set of equivalent
constraints, which can be
treated as some coordinates in the manifold,
is known.
The question is how, in a  constructive way  to find this transformation
matrix.
According to (\ref{eq:abconstr}), the matrix \({\cal D}_{\alpha \beta}  \)
must satisfy the set of the {\it nonlinear} differential equations
\bea                \label{eq:non}
    \{ {\cal D}_{\alpha \gamma} (p, g)\varphi_\gamma (p,q)  ,
{\cal D}_{\beta \sigma} (p, g)\varphi_\sigma  (p,q)
\}\, &  = &\, 0.
\eea
The  statement of abelianization means a possibility of
finding  a particular solution for  these very complete
{\it nonlinear} differential equations.
Beyond  question  eq. (\ref{eq:non}) in this form is not of
any practical value; but it has been  shown \cite{Gogil1} that there is a
particular solution to this equation  and it can be  represented as
\be                \label{eq:mattran}
{\cal D} \, = \underbrace{{\cal D}^1 (p, q)\cdots {\cal D}^m (p, q)}_{m}
 \ee
where each matrix \({\cal D}^k \) is again  represented in product form of
\(k \)'s  \( m \times m \)  matrices
\be                \label{eq:matrix}
{\cal D}^k \, = {\cal R}^{a_k +k} (p, q)\prod_{i=k-1}^{0}{\cal S}^{a_k +i}
(p, q)
\ee
(\(a_k \equiv k(k+1)/2 ) \) and
\bea \label{eq:r}
\begin{array}{lr}
{}~~~~~~~~~~~~~~\overbrace{~~~~~~~~~~~~~~}^{k}\overbrace{~~~~~~~~~~~~~~~~~~~~}^{m-k} \\
{\cal R}^{a_k+k} = \left (
\begin{array}{cc}
{\fbox{\raisebox{0.mm}[10mm][5mm]{ \quad {\mbox{\huge I}} \quad}}}  &
\mbox{\huge 0}      \\
\mbox{\huge 0}     &
\fbox{\raisebox{0.mm}[10mm][5mm]{ \quad ${\mbox{\huge B}}^{a_k+k}$ \quad}}
\end{array}
\right )
\end{array}
\eea
\bea \label{eq:s}
\begin{array}{lr}
{}~~~~~~~~~~~~\overbrace{~~~~~~~~~~~~~~~~~~~~~~~~~~~~}^{k}
\overbrace{~~~~~~~~~~~~~~~~~~~~~~~~~~}^{m-k} \\
{\cal S}^{a_k+i} =
\left (
\begin{array}{ccccc|ccccc}
1       &   0      & \cdots  &   0      &   0      &    0       &    0
&\cdots  &   0     &    0   \\
0       &   1      & \cdots  &   1      &   0      &    0       &    0
&\cdots  &   0     &    0   \\
\vdots  &  \vdots  & \ddots  &  \vdots  &  \vdots  &  \vdots & \vdots   &\ddots
  &  \vdots &  \vdots  \\
0       &  0       & \cdots  &   1      &   0      &    0      &    0
&\cdots   &   0     &    0    \\
0       &  0       & \cdots  &   0      &   1      &    0     &    0
&\cdots   &   0     &    0    \\
\hline
0       &  \cdots  &  C_{k+1}^{a_k+i}    &   \cdots  &   0      &      1     &
 0     & \cdots  &   0     &   0    \\
0       &  \cdots  &  C_{k+2}^{a_k+i}    &   \cdots  &   0      &      0     &
 1     & \cdots  &   0     &   0    \\
\vdots  &  \vdots  & \vdots          &   \vdots  &  \vdots  &  \vdots & \vdots
& \ddots  & \vdots  &  \vdots \\
0       &  \vdots  & C_{m-1}^{a_k+i} &   \cdots  &   0      &     0      &  0
   &  \cdots          &    1    &   0   \\
0       &  \cdots  &  C_m^{a_k+i}    &   \cdots  &   0      &     0      &  0
   & ~~~\cdots ~~~ &    0    &   1
\end{array}
\right ) \\
{}~~~~~~~~~~~~\underbrace{~~~~~~~~~~~~~~~}_{k-i} &
\end{array}
\eea
and satisfies a set of {\it linear } differential equations ( see below (
\ref{eq:lin1}), (\ref{eq:lin2})).  Just the linear character of these
equations allows one to speak about a practical use of the proposed method
of abelianization.
The constraints
which are obtained as a result of the action of \( k\)'s matrices
(constraints at the \(a_k + k\) -th step )
\be
\Phi_\alpha^{a_k+k} =\left (\underbrace{{\cal D}^k \cdot{\cal D}^{k-1}
 \cdots {\cal D}^k}_{m}\right )_{\alpha \beta}  \Phi_\beta^{0}
\ee
obey the algebra where \(k\) constraints have   zero Poisson brackets
with any one.
 From the algebraic standpoint  this method of  abelianization represents
an  iterative procedure of constructing of ``equivalent''
algebras \({\cal A}^{a_i}\) of constraints \( \Phi_\alpha^{a_i} \)

\be \label{eq:setalg}
\framebox[140mm]{ \raisebox{0ex}[4ex][3ex]{
$ {\cal A}^{0}\underbrace{\stackrel{{\cal S}^{1}}{\to}{\cal A}^{1}
\stackrel{{\cal R}^{2}}{\to}}_{{\cal D}^1}{\cal A}^{2}
\underbrace{\stackrel{{\cal S}^{3}}{\to}
{\cal A}^{3}\stackrel{{\cal S}^{4}}{\to}{\cal A}^{4}
\stackrel{{\cal R}^{5}}{\to}}_{{\cal D}^2}{\cal A}^{5}
\dots \underbrace{\stackrel{{\cal S}^{a_k}}{\to}
{\cal A}^{a_k}\dots
\stackrel{{\cal R}^{a_k+k }}
{\to}}_{{\cal D}^k}{\cal A}^{a_k+k} \dots
$}}
\ee
The abelianization procedure consists in  \(a_m \)'s  steps
for construction of the \(m \)  - dimensional abelian algebra
equivalent to the initial non - Abelian one
in such a manner that  at the \(a_k\) - th step
the obtained algebra \( {\cal A}^{a_k} \) possesses a center
with \( k\) elements
\( {\cal Z}_k [ A ] = \left( \Phi_1^{a_k}, \Phi_2^{a_k}, \dots
\Phi_k^{a_k} \right) \)
The matrix \( {\cal D}^k\) converts the algebra \({\cal A}^{k}\)
to the algebra \({\cal A}^{k+1}\) in which the center contains
one element more than previous.

The validity  of the representation  (\ref{eq:matrix})
with the matrices \({\cal S}\) and \( {\cal R} \)
was proved in \cite{Gogil1} by induction.
It has been shown  that if
\( \Phi_\alpha^{a_k} \) - are   constraints
(obtained as result of
action of the \(k-1 \) matrices \( {\cal D}^i\) )
with  the algebra having the  center
\( {\cal Z}_k [ A ] = \left( \Phi_1^{a_k}, \Phi_2^{a_k}, \dots ,
\Phi_k^{a_k} \right) \),
then  a matrix
\( {\cal D}^k\)  from (\ref{eq:mattran}) performs the transformation
to the  new constraints
\be
\Phi_\alpha^{a_{k+1}-1} ={{\cal D}^k }_{\alpha \beta } \Phi_\beta^{a_k+1}
\ee
which form  the algebra with  the center
{}~\({\cal Z}_{k+1} [ A ] ~ = ~ \left( \Phi_1^{a_{k+1}}, \Phi_2^{a_{k+1}} ,
\dots,
\Phi_k^{a_{k+1}}, \right.\\
\left.\Phi_{k+1}^{a_{k+1}} \right) \)
if the matrices  \({\cal S}\) and \( {\cal R} \)
are the solutions to the following set of  linear differential
equations
\bea \label{eq:lin1}
\left.
\begin{array}{rcr}
\{ \Phi_{1}^{a_k+i-1}, S_{\alpha_k}^{a_k+i} \}& =& 0  \\
\vdots~~~~~~~~~~~~~~~~~~\vdots  &~~~ &\vdots \\
\{ \Phi_{k-1}^{a_k+i-1}, S_{\alpha_k}^{a_k+i} \}& =& 0 \\
\end{array}\right\}\;\;\; \Longrightarrow
\{ \Phi^{a_k+i-1}_{{\overline{\alpha}}_k}, S_{\alpha_k}^{a_k+i} \} = 0
\eea

\bea \label{eq:lin2}
\{ \Phi_k^{a_k+i-1}, S_{\alpha_k}^{a_k+i} \}& = &
f^{a_k+i-1}_{k \alpha_k \gamma_k}
  S_{\gamma_k}^{a_k+i}  - f^{a_k+i-1}_{k \alpha_k i+1}
\eea

\bea
\left.
\begin{array}{rcr}
\{ \Phi_{1}^{a_k+k-1}, B_{\alpha_k\beta_k}^{a_k+k}  \}& =& 0  \\
\vdots~~~~~~~~~~~~~~~~~~\vdots  &~~~ &\vdots \\
\{ \Phi_{k-1}^{a_k+k-1}, B_{\alpha_k\beta_k}^{a_k+k} \}& =& 0 \\
\end{array}\right\}\;\;\; \Longrightarrow
\{ \Phi^{a_k+k-1}_{{\overline{\alpha}}_k}, B_{\alpha_k\beta_k}^{a_k+k} \} = 0
\eea

\bea
\{ \Phi_k^{a_k+k-1}, B_{\alpha_k\beta_k}^{a_k+k} \}  =
- f^{a_k+k-1}_{k  \gamma_k \beta_k}B^{a_k+k}_{\alpha_k \gamma_k}
\eea
where
\(\alpha_k = k+1,\dots, m \;,{\overline{\alpha}}_k = 1,2,\dots, k-1 \)
and  \(f^{a_k+i}_{\alpha  \gamma \beta}\) are the
structure functions of the constraints algebra \(A^{a_k+i}\)  at the
\({a_k+i}\) -th step.

\begin{center}
{\it i.i.i. Construction of physical coordinates  via
Dirac's transformation }
\end{center}

However, for our purpose, to construct the coordinates of the physical
subspace we can act in a slightly different  way.
The proposed method of explicit realization of the reduced phase space
consists on  the step by step elimination of
 ignorable
coordinates of the phase space with the help of construction
of the corresponding  Abelian subalgebra.
This can be achieved at three steps
\begin{itemize}
\item[a)] first, we obtain an equivalent  to the initial algebra with
central element \(\varphi_1\)
\[
    \{ \varphi_1 (p,q),  \Phi_\beta  (p,q)\}\,   = \, 0.
\]
\item[b)] next, perform the canonical transformation to a new set of
 coordinates so that
\[
\overline{P}_1
 = \varphi_1 \left ( q_i , p_i \right ),
\quad \{\overline{Q}_1 , \overline{P}_1\}
 = 1
\]
\item[c)] last, restrict to the \(2n - 1\)-dimensional
submanifold spanned by the coordinates \(
\overline{Q}_{\alpha_1} , \overline{P}_{\alpha_1} ,
\)
and to the algebra \(
\overline{\Phi}_{\alpha_1} \equiv {\Phi_{\alpha_1}}
\Bigl\vert_{  \overline{P}_1 = 0}
\)
\end{itemize}

\noindent{\underline{\it \(a)\) First step  }}
For determination of a new algebra with one central  constraint
\( \varphi_1\) one can act as it was described in \cite{Gogil1} :
\begin{itemize}
\item  exclude  \( \varphi_1\)
>from the left hand side of eq. (\ref{eq:non});
\item then  realize abelianization with all others
\end{itemize}
Tor achieve the first, one can perform the transformation with the
matrix \({\cal S}^1 \)
\[
\Phi^1_\alpha = {\cal S}^1_{\alpha \beta}\varphi_\beta
\]
of type (\ref{eq:s})
\be
{\cal S}^1 =  \left (
\begin{array}{ccccc}
1        &   0      &   0    & \cdots  &   0    \\
C_2      &   1      &   0    & \cdots  &   0    \\
C_3      &   0      &   1    & \cdots  &   0     \\
\vdots   &   \vdots & \vdots & \ddots  & \vdots  \\
C_m      &   0      &   0    &  \cdots  & 1
\end{array}
\right )
\ee
or in the expanding form
\bea
\Phi^1_1 &= &\Phi^0_1 =\varphi_1 \nn\\
\Phi^1_{\alpha_1}& =& \varphi_{\alpha_1} +
C^1_{\alpha_1}\varphi_1
\eea
The new constraints algebra remains the algebra of first class
\bea
\{\Phi^1_{1},  \Phi^1_{\alpha_1}  \}&  = &
f^1_{1 \alpha_1 1}  \Phi^1_1 +
f^1_{1 {\alpha_1} \gamma_1}  \Phi^1_{\gamma_1}\nn\\
\{\Phi^1_{\alpha_1},  \Phi^1_{\beta_1}  \}&  = &
f^1_{ {\alpha_1} {\beta_1} {\gamma_1}}  \Phi^1_{\gamma_1}
+ f^1_{ \alpha_1 \beta_1 \gamma_1}  \Phi^1_{\gamma_1}
\eea
and the new structure functions \(f^1_{ \alpha \beta \gamma} \)
are determined via the old one  \(f^1_{ \alpha \beta \gamma} \) and
the transformation functions \(C^1_{\alpha_1}\) as follows
\bea \label{eq:str1}
f^1_{1 {\alpha_1} 1}& = &f_{1 {\alpha_1} 1} + f_{1 {\alpha_1} \gamma_1}
C^1_{\gamma_1} + \{\Phi^0_{1},  C^1_{\alpha_1}  \}\\
f^1_{\alpha_1 {\beta_1} 1}& = &{{1}\over{2}} \left(\;
f_{{\alpha_1}{\beta_1} 1} -
 f_{{\alpha_1}{\beta_1} \gamma_1}C^1_{\gamma_1}
+ \{ C^1_{\alpha_1}, C^1_{\beta_1}\}\Phi_1^0 \;\right)  - \nn\\
&-& f^1_{1 {\alpha_1} 1} C^1_{\beta_1} +
\{\Phi^0_{\alpha_1}, C^1_{\beta_1} \}
-(\alpha_1 \leftrightarrow \beta_1 ) \\
f^1_{{\alpha_1}\beta_1 \gamma_1}& =& f_{{\alpha_1}\beta_1 \gamma_1}
+ C^1_{\alpha_1}f_{1 {\beta_1} \gamma_1 } -
C^1_{\beta_1}f_{1 {\alpha_1} \gamma_1 } \\
f^1_{1 {\alpha_1} \gamma_1}& =& f_{1 {\alpha_1} \gamma_1}
\eea
One can now choose the transformation functions \(C^1_{\beta_1}\)
so that the Poisson bracket of first constraints \( \Phi_1^1 \) with
all other modified constraints do not contain it
\bea
\{ \Phi^1_1 (p,q),  \Phi^1_{\alpha_1} (p,q)\}\,&  = &\,
\sum_{\gamma \not= 1} f^1_{ 1 \alpha_1 \gamma}  (p,q) \Phi^1_\gamma  (p,q).
\eea
these \(m-1 \) requirements : \( f^1_{1 {\alpha_1} 1}  = 0  \)
according to eq. (\ref{eq:str1} ) means that the transformation function
must satisfy  the
following set of linear nonhomogeneous differential equations
\bea \label{eq:1}
\{\Phi^0_{1},  C^1_{\alpha_1} \} =
- f_{1 {\alpha_1} 1} + f_{1 {\alpha_1} \gamma_1}
C^1_{\gamma_1}
\eea
Note that the problem of  existence of solution to
such a set of equations is studied very well
( see e.g. \cite{Kur} )
Suppose, we find some particular solution \( C^1_{\alpha_1}\) to
(\ref{eq:1}), then one can determine all structure functions
of the modified algebra according to eq.(\ref{eq:str1}):
\bea \label{eq:str11}
f^1_{1 {\alpha_1} 1} & = & 0 \\
f^1_{\alpha_1 {\beta_1} 1}& = &
f_{{\alpha_1}{\beta_1} 1} -
 f_{{\alpha_1}{\beta_1} \gamma_1}C^1_{\gamma_1}
+ \{ C^1_{\alpha_1}, C^1_{\beta_1}\}\Phi_1^0\nn\\
&+& \{\Phi^0_{\alpha_1}, C^1_{\beta_1} \}
+\{\Phi^0_{\beta_1}, C^1_{\alpha_1} \} \\
f^1_{{\alpha_1}\beta_1 \gamma_1}& =& f_{{\alpha_1}\beta_1 \gamma_1}
+ C^1_{\alpha_1}f_{1 {\beta_1} \gamma_1 } -
C^1_{\beta_1}f_{1 {\alpha_1} \gamma_1 } \\
f^1_{1 {\alpha_1} \gamma_1}& =& f_{1 {\alpha_1} \gamma_1}
\eea

Now let us again keep first constraint unchanged and
perform  the Dirac transformation
on the remaining  part of the constraints
\( \Phi_{\alpha_1},\;\; \alpha_1 = 2,3, \dots, m \)
\bea
\Phi^2_1 &= & \Phi^1_1 = \Phi^0_1 =\varphi_1 \nn\\
\Phi^2_{\alpha_1}& = & B^2_{\alpha_1 \beta_1}
\Phi^1_{\beta_1}
\eea
with the requirement that new constraints have zero Poisson
brackets with the first one \(\Phi^1_1 \)
\bea
    \{ \Phi^2_1 ,  \Phi^2_{\alpha_1}\}\, &  = &\, 0.
\eea
One can verify that this requirement means that the transformation
functions \(B_{\alpha_1 \beta_1} \) are the solutions to the equation
\bea \label{eq:2}
\{\Phi^1_{1},  B^2_{\alpha_1\beta_1} \} =
- f_{1 {\gamma_1} \beta_1}B^2_{\alpha_1\gamma_1}
\eea
With  the help of a solution of eq. (\ref{eq:2})  the modified
algebra has the
following constraints:
\bea \label{eq:str2}
f^2_{1 {\alpha_1} 1} & = & 0 \\
f^2_{\alpha_1\beta_1 1}&=& B^2_{\alpha_1\delta_1}B^2_{\beta_1 \sigma_1}
f^1_{\delta_1 \sigma_1 1 } \\
f^2_{\alpha_1 {\beta_1} \gamma_1}& = &
\left[ \{ B^2_{\alpha_1 \delta_1}, B^2_{\beta_1 \sigma_1}\}
\Phi_1^{\sigma_1}
+
\{ B^2_{\alpha_1 \delta_1}, \Phi^1_{\sigma_1}\}
B^2_{\beta_1 \sigma_1}\right. - \nn\\
&-&\left.
\{ B^2_{\beta_1 \delta_1}, \Phi^1_{\sigma_1}\}
B^2_{\alpha_1 \sigma_1}
+
B^2_{\alpha_1\kappa_1} B^2_{\beta_1 \sigma_1}f^1_{\kappa_1\sigma_1\delta_1 }
\right](B^2)^{-1}_{\delta_1 \rho_1}
\eea

Thus as a result of two transformations
\({\cal D}^{1} = {\cal S}^{1}{\cal R}^{2} \) we obtain the  modified
algebra \( {\cal A}^2 \) of constraints
\( \Phi_\alpha^2 \) with the central element   \( \Phi^2_1 \)
\bea
\{\Phi^2_{1},  \Phi^2_{\alpha_1}  \}&  = & 0 \\
\{\Phi^2_{\alpha_1},  \Phi^2_{\beta_1} \}&  = &
f^2_{ {\alpha_1} {\beta_1} 1}  \Phi^2_{1}
+
f^2_{ \alpha_1 \beta_1 \gamma_1}  \Phi^2_{\gamma_1}
\eea
It  is  to be noted  that  due to
central element nature of  \( \Phi^2_1 \),
the structure functions obey the following
property:
\bea \label{eq:intt1}
\{\Phi^2_{1},  f^2_{\alpha_1 \beta_1 \gamma}  \}&  = & 0
\eea

So, with the help of two Dirac' s transformations we obtain an equivalent
to  the  initial algebra with one central element
\bea
    \{ \varphi_1 (p,q),  \Phi_{\alpha_1}(p,q)\}\, &  = &\, 0.
\eea

\noindent\underline{\it b) Second step }
Now one can note that as a result of two transformations
the first constraint `` commutes '' with all others but it can arise
with  left hand side of
\bea
\{ \Phi_{\alpha_1} (p,q),  \Phi_\beta  (p,q)\}\, &  = &\,
c_{ \alpha_1 \beta_1 1}  (p,q) \varphi_1  (p,q) + \nn\\
 & + & c_{ \alpha_1\beta_1 \gamma_1}  (p,q) \varphi_{\gamma_1}  (p,q).
\eea
How  can one  shake off this term ?
The following observation can help us.
One can always pass to a new canonical coordinate
\bea
q_i  & \mapsto & Q_i = Q_i \left (  q , p \right )
\nn\\
p_i & \mapsto & P_i = P_i \left ( q , p \right ),
\eea
so that one of the new momentum
will be equal to the first constraint \( \varphi_1 \)
\bea
\overline
{P}_1 & = &\varphi_1 \left (  q_i , p_i \right )
\eea
In these new canonical coordinates  eq.
(\ref{eq:intt1})
means that the new structure functions \( f^2_{\alpha_1 \beta_1 \gamma}\)
do not depend on the coordinate \(\overline{Q}_1 \)
\bea \label{eq:ind}
\{\overline{P}_{1},  f^2_{\alpha_1 \beta_1 \gamma} (P,Q) \}  =  0
\quad
\mapsto
\quad
\frac{\partial f^2_{\alpha_1 \beta_1 \gamma}}{\partial
\overline{Q}} = 0
\eea
\noindent\underline{\it c) Third step }
Let us now consider the  new set of constraints
obtained as follows:
\be
\overline{\Phi}_{\alpha_1} \equiv {\Phi_{\alpha_1}}
\Bigl\vert_{  \overline{P}_1 = 0}
\ee
It is not worth noting that this transition to the new set of constraints
\(\overline{\Phi}_{\alpha_1} \equiv {\Phi_{\alpha_1}}
\Bigl\vert_{  \overline{P}_1 = 0}
\)
is again the Dirac transformation of type (\ref{eq:Diractran})
with the matrix
\be
 C_\alpha = \sum_{k=1}^{\infty}
\frac{\partial^{k}
 \Phi^2_{\alpha}}{\partial^{k}
  P_1} \Bigl\vert_{  \overline{P}_1 = 0}
\overline{P}_1^
{k}
\ee
The algebra of
new constraints \( \Phi_{\alpha_1}\) has a closed form
(the right -  hand side do not depend on \( P_1 \) with
 the structure functions
\be
\overline{
 f}^2_{\alpha_1 \beta_1 \gamma_1} =
 f^2_{\alpha_1 \beta_1 \gamma_1}
 \Bigl\vert_{  \overline{P}_1 = 0}
\ee
depending only on the remaining part of the coordinates
\(
\overline{Q}_{\alpha_1} , \overline{P}_{\alpha_1},
\)
due to the  property  (\ref{eq:ind})
Thus, by these admissible manipulations we reduce our problem to
the  equivalent  one  only  for \(m- 1\) - dimensional algebra of
constraints in \(2n-2\) - dimensional phase space.
 We will  obtain the desired physical coordinates,
by acting  in such a manner step by step.


\section{\sf \,\, Conditions for admissible gauges}


\bigskip
As it has been mentioned above the  generalized canonical transformations
{}~\cite{Bergman}
are the those  preserving the form of all constraints
of the  theory as well as the canonical form of the equations of motion .
Thus all forms of representation of the singular theory must be
connected with each other by this  a kind of transformations.
This allows us to give the  following definition of an
admissible gauge :
\begin{quote}
{\it A gauge is admissible if and only if there is  a
\underline{generalized} canonical equivalence
between the reduced phase space obtained by the gauge fixing
method and the gaugeless one.}
\end{quote}
Sufficient condition for a gauge to be  admissible
consists in dependence of gauge fixing conditions only on nonphysical
variables.
The above described method  allows us to find a sufficient condition on the
gauge fixing
functions to belong to the class of canonical gauges for  which
the equivalence between  the gauge fixing method and  gaugeless one is
fulfilled.

How can one  recognize the existence of such an equivalence
and what is the necessary and sufficient condition for a gauge to
belong to a class of  admissible gauges.

To get the answer, we must study, the general structure of the reduced
theory.
Let us again return to the case of the Abelian theory or to the
 non - Abelian theory rewritten in the  equivalent abelian form.
Having represented the  theory in to the form
(\ref{eq:1str}) where two sectors, physical and nonphysical,
are separated, it is clear that the most suitable gauge conditions are
functions depending only on the ignorable coordinates
\(\chi_\alpha \equiv \chi_\alpha (\overline{Q} ) \)
\bea \label{eq:cangauge}
\{ \chi_\alpha , {Q}^{\ast}_i \} & = & 0, \nn\\
\{ \chi_\alpha, {P}^{\ast}_i \}  & = & 0
\eea
Gauges of this type will be called the {\it canonical gauges}.
Now the question is how to reformulate  this property of
independence on the physical variables  in the
initial coordinates \(p, q \).
At this point it is convinient to use the fact that
in virtue of the definition of the physical Hamiltonian,
(\ref{eq:physham})
the requirement of independence of a  gauge on physical
variables can be written
in canonically invariant form as
\footnote{Certainly there is a possibility when some of the physical
coordinates \( {Q}^{\ast} \) do not enter in to the physical Hamiltonian
due to the some global symmetry and thus  are usual ignorable
coordinates. As a result
they break through this requirement and can present in the gauge
condition \( \chi \), but owing to their ignorable character
they can be treated in same manner  as gauge noninvariant
`` ignorable coordinates '' \(\overline{Q} \).}
\be\label{eq:1cond}
\{ \chi_\beta (p, q), H_{Ph}(p,q)\}
\Bigl\vert_{\Gamma^\ast} = 0
\ee
In this form this condition  is far from practical usage.
However, one could transform it to a very simple form with  the
help of the Dirac bracket
Indeed, in the special coordinates \( Q,P \)
starting from the representation
(\ref{eq:hamrep})  one can extract from \(\overline{H}_0 \)
\(\overline{P}_\alpha \) - independent physical Hamiltonian and
write down  the following decomposition for
the canonical Hamiltonian
\bea\label{eq:dec}
\overline{H}_C (P,Q)& = & \overline{H}_0({Q}^{\ast}, {P}^{\ast},  \overline{P})
 + \overline{\Psi}_\alpha (Q,P) \overline{P}_\alpha = \nn\\
&=&
\overline{H}_{Ph}(P^\ast, Q^\ast ) +  F_\alpha (Q,P) \overline{P}_\alpha
\eea
with functions \( F_\alpha \) determined by $H_0$ and $\Psi$ .
Now, taking into account that nor the canonical
gauge  nor the matrix
\(\overline{\Delta}_{\alpha \beta}  =
\{\chi_\alpha, \overline{P}_\beta \} \)
depend  on the physical variables ,
we have
\bea
\{ \overline{\chi}_\alpha (\overline{Q}), \overline{H}_C(P,Q)\}
&=& \overline{\Delta}_{\alpha \beta }(\overline{Q}) F_\beta(P,Q) +
\{ \overline{\chi}_\alpha(\overline{Q}), F_\beta(P,Q)\}
\overline{P}_\beta \label{eg:1}  \\
\{ \overline{\Delta}_{\alpha \beta } (\overline{Q}), \overline{H}_C(P,Q)\}
&=&\{ \overline{\Delta}_{\alpha \beta}(\overline{Q}) F_\gamma(P,Q)\}
\overline{P}_\gamma
 +
\{ \overline{\Delta}_{\alpha \beta }(\overline{Q}),
\overline{P}_\gamma \}
F_\gamma(P,Q)\nn
\eea
Assuming for the  moment that
\(\{ \overline{\Delta}_{\alpha \beta }(\overline{Q}),
\overline{P}_\gamma \} \not= 0\)
and excluding from (\ref{eg:1}) functions
\(F_\gamma(P,Q) \)
one get
\be \label{eq:11suffcon}
\{ \overline{\Delta}_{\alpha \beta} (\overline{Q}) , \overline{H}_C(P,Q) \}
\Bigl\vert_{ \Gamma^\ast} =
\{ \overline{\Delta}_{\alpha \beta }(\overline{Q}),
\overline{P}_\gamma \}
\overline{\Delta}^{-1}_{\gamma \sigma}(\overline{Q})
\{ \overline{\chi}_{\sigma} (\overline{Q}) , \overline{H}_C(P,Q) \}
\Bigl\vert_{ \Gamma^\ast}
\ee
Taking into account
the definition (\ref{eq:PDB}) of the Dirac bracket
this conditions can be rewritten in a more attractive form
\be \label{eq:0suffcon}
\{ \overline\Delta_{\alpha \beta} (\overline{Q}) , \overline{H}_C(P,Q) \}
{}_{ D( \overline{P}, \overline{\chi})}\;
\Bigl\vert_{ \overline{P} = 0, \: \overline{\chi} = 0 }\, = \,0
\ee
Return to the case when
\(\{ \overline{\Delta}_{\alpha \beta }(\overline{Q}),
\overline{P}_\gamma \} = 0\) we see that there is possibility
when (\ref{eq:0suffcon}) is satisfied by the gauge condition
depending on the some physical coordinates
\[
{\overline{\chi}}_\alpha  =  \overline{Q}_\alpha +
f_\alpha({Q}^\ast)
\]
But as it was mentioned above this dependence is not a significant,
by canonical transformation one can get rid of it.

Now let us try to rewrite  the condition  (\ref{eq:0suffcon})
in to the old coordinates \(p,q\) and for the non-Abelian
form of constraints
\be \label{eq:abec}
\varphi_\alpha =
{\cal D}_{\alpha \beta}\overline{P}_\beta
\ee
Due to the well - known observation  \cite{Bergman}, \cite{Sudarsh}
the Dirac bracket is generalized
canonical invariant object
\[
\{ \overline{F}(P,Q), \overline{G}(P,Q) \}
{}_{D(\overline{P}, \overline{\chi}) }=
\{ {F}(p,q),{G}(p,q) \}
{}_{D(\varphi, \chi)}
\]
and thus instead of
(\ref{eq:11suffcon}) one can write down

\be \label{eq:suffcon}
{\cal D}_{\alpha \gamma}
\{\Delta_{\gamma \beta}(p.q), {H}_C(p,q)\}
{}_{D(\varphi, {\chi})}
+
\Delta_{\gamma \beta} (p.q)
\{{\cal D}_{\alpha \gamma}, {H}_C(p,q)\}
{}_{D(\varphi, {\chi})}\;
\Bigl\vert_{ \varphi= 0,{\chi} = 0 } = 0
\ee
Now let us prove that the matrix of abelianization depends only
on the variables \( \overline{P} \) and \( \overline{Q} \)
and thus
\be
\{{\cal D}_{\alpha \gamma}  , {H}_C(p,q) \}
{}_{D(\varphi, {\chi})}\;
\Bigl\vert_{ \varphi= 0, \:{\chi} = 0 }\, = \,0
\ee
It can be verified as follows.
As it is known \cite{DiracL}, \cite{Gen}
the generator of gauge transformations
can be represented as a sum of first class constraints
in non - Abelian form
\[
G = \varepsilon_\alpha(q,p,t) \varphi_\alpha (q,p).
\]
or with the help of abelian one
\be \label{eq:genab}
G=\overline{{\varepsilon}}_\alpha(\overline{Q}, \overline{P}, t)
 \overline{P}_\alpha.
\ee
It is necessary to note that in eq. (\ref{eq:genab})
the parameters of gauge transformations
\({\varepsilon}_A(\bar Q,\) depend only
the ignorable coordinates
\( \overline Q, \overline P \) in virtue of separable
form of eqs. (\ref{eq:heq}).
According to the eq.(\ref{eq:genab}) any gauge invariant
function \( I \)
\be \label{eq:abe}
\{I, G\} = 0
\ee
depends only on the variables  $ Q^\ast $ and $ P^\ast $.
The  $ Q^\ast, P^\ast $ compose the basis of
gauge invariant variables.
Therefore  from eq. (\ref{eq:abe})
with generator \(G\)
expressed via non - Abelian constraints and matrix of abelianization
according to the (\ref{eq:abec})
\[
G = \bar{\varepsilon}_\alpha(\overline{Q},\overline{P},t){\cal
D}_{\alpha \beta}^{-1}
{\varphi}_\beta.
\]
one can get for \( I \equiv
Q^\ast, P^\ast \)

\bea
&& \{Q_i^\ast, G\} = 0 \quad  \Longrightarrow  \quad
 \frac{\partial{\cal D}_{\alpha \beta}^{-1}}{\partial Q_i^\ast}=0\nn\\
&& \{P_i^\ast,G\}=0 \quad   \Longrightarrow   \quad
 \frac{\partial{\cal D}_{\alpha \beta}^{-1}}{\partial P_i^\ast}=0.
\eea
where the functional independence of constraints and
nonsingularity  of matrix ${\cal D}$, have been exploited.

This completes the proof of independence of the matrix of abelianization
on the variables \( {P}^\ast\) and \( {Q}^\ast \)
and thus finally we get the desired condition
\be \label{eq:suffcond}
\{ \Delta_{\alpha \beta} (p,q) , H_C(p,q) \}{}{}_{D}\;
\Bigl\vert_{ \varphi = 0, \:\chi = 0 }\, = \,0
\ee
where   the matrix
\( \Delta_{\alpha \beta } =
\{\chi_\alpha, \varphi_\beta \} \)
is calculated with  the non - Abelian constraints.

\section{\sf Christ and Lee model}

\bigskip
\subsection{\sf Abelian Christ \& Lee model}

\begin{center}
{\it i. Reduction without gauge fixing}
\end{center}

Let us consider a simple  mechanical system for which
one can explicitly realize the above described scheme of construction
of true dynamical degrees of freedom : Christ and Lee model \cite{Christ}
\be \label{eq:CLl}
{\cal L}=\frac{1}{2}\bigl(\dot x_1^2+\dot x_2^2+y^2(x_1^2+x_2^2)\bigr)-
y(\dot x_1x_2-x_1\dot x_2)-V(x_1^2+x_2^2),
\ee
where  \((x_1, x_2, y) \) are independent coordinates .
The rank of the Hessian matrix is equal to two and thus we have one primary
constraint
\be  \label{eq:prim1}
\varphi_1^1 = p_y = 0,
\ee
according to the definition of a canonical momentum $p_y$.
The corresponding total Hamiltonian is
\be
H_T=\frac{1}{2}(p_1^2+p_2^2)-y(x_1p_2-x_2p_1)+V(x_1^2+x_2^2)+u(t)p_y.
\ee
 From the stationarity condition of primary constraint (\ref{eq:prim1})
we get the secondary constraint
\be
\varphi_1^2 = x_1p_2 - x_2p_1 = 0,
\ee
It  is easy to verify that ternary constraints are absent
and that the  constraints  $\varphi_1^1$  and  $\varphi_1^2$
are the first class ones
\[ \{\varphi_1^1,\varphi_1^2\}=\{p_y,x_1p_2-x_2p_1\}=0
\]
This means that there are  gauge transformations
generated  by
\be \label{eq:ggen}
G=-\dot{\varepsilon}(t)p_y + \varepsilon(t)(x_1p_2-x_2p_1).
\ee
These gauge transformation
are nothing else but a rotation around the
axis orthogonal to the plane  $(x_1,x_2)$ on the angle  $\varepsilon(t)$.
\bea
x'_1=x_1+\{x_1,G\}=x_1-\varepsilon(t)x_2 \nn \\
x'_2=x_2+\{x_2,G\}=x_2+\varepsilon(t)x_1 \nn \\
y'=y+\{y,G\}=y-\dot \varepsilon(t)
\eea
Now let us  introduce  the Levi - Civitta transformation to a special
set of canonical coordinates
\((y, p_y), (x_1, p_1), (x_2, p_2) \mapsto  (Y, P_Y),
(R, P_R), (\overline{\Theta},
\overline{P}_{\overline{\Theta}}) \)
so that the new  momentum \(\overline{P}_{\overline{\Theta}} \)
is equal to the secondary constraint \( \varphi_1^2 \)
\bea \label{eq:lctr}
Y = y ,                        & & P_Y = p_y ,\\
R = \sqrt{x_1^2 + x_2^2} \,, & & P_R = \frac{ x_1 p_1 + x_2 p_2}
{\sqrt{x_1^2 + x_2^2}} \,, \\
\overline{\Theta} = \arctan\left(\frac{x_2}{x_1}\right)\, , &  &
\overline{P}_{\overline{\Theta}} = x_1 p_2 - x_2 p_1.
\eea
These transformations  are canonical and  non - singular
with the inverse
\bea
y = Y  ,                        & & P_Y = p_y ,\\
x_1 = R\cos \overline{\Theta} , & & p_1 =
P_R\cos \overline{\Theta}  -
\frac{\overline{P}_{\overline{\Theta}}}{R} \sin \overline{\Theta} \,,\\
x_2 = R\sin \overline{\Theta}\,, & & p_2 =
P_R\sin\overline{\Theta} +
\frac{\overline{P}_{\overline{\Theta}}}{R} \cos \overline{\Theta}\,.
\eea
everywhere except one point \( R = 0 \) if we suppose that
\( 0 < \overline{\Theta} < 2 \pi \).
In terms of these variables the total Hamiltonian has the form
\bea
H_T = \frac{1}{2}(P_R^2+
\frac{\overline{P}^2_{\overline{\Theta}}}{R^2}) -
Y \overline{P}_{\overline{\Theta}} + V(R^2) + u_Y P_Y .
\eea
Note that this form is in accordance with the general representation
(\ref{eq:hamrep}) with the physical Hamiltonian
\be
H_{Ph} =\frac{1}{2}P_R^2 + V(R^2)
\ee
and the function \(\Psi \)
\bea
\Psi =   \left (\frac{\overline{P}_{\overline{\Theta}}}{2R^2} -
Y \right)\overline{P}_{\overline{\Theta}}
\eea
And finally from the equations of motions
\bea \label{eq:mheq}
&& \dot{R}\,\,  =  P_R\,, \nn\\
&& \dot{P}_R  = -\frac{\partial V(R^2)}{\partial R }\,,\nn\\
&& \dot{\overline{P}_{\overline\Theta}} =  0, \quad
 \dot{\overline{\Theta}}_\alpha  =  \overline{u}_\Theta(t)\nn\\
&& \dot{P}_{Y}  = 0, \quad \dot{Y}  = \overline{u}_Y(t)
\eea
we  conclude that the transformation (\ref{eq:lctr}) alows to separate the
phase space  coordinates into two groups: gauge invariant $R, P_R,
P_Y, \overline{P}_{\overline\Theta}$ and  noninvariant
ignorable coordinates $ Y, \overline{\Theta}$.
This means that we achieve reduction, and now it is enough to
pass to the constraint shell  (in this case, it means
that we must put
constraints  $P_Y$ and $\overline{P}_{\overline\Theta}$ equal to zero).
Thus, we get the elimination of nonphysical variables without
gauge fixing only through
passing  to the constraint shell.

\bigskip

\begin{center}
{\it i.i. Gauge fixing method : example of a nonadmissible gauge }
\end{center}

\bigskip

Now we can return to a gauge fixing scheme .
Any correct reduction via gauge fixing of the  considered  constrained
system must lead to the theory that is canonically equivalent to it.
First, let us write down the canonical gauge for the system
(\ref{eq:CLl}).
\bea \label{eq:cclg}
&&\chi_1 \, \equiv \, y = 0, \nn \\
&&\chi_2 \,\equiv \, \arctan\left(\frac{x_2}{x_1}\right)  = constant.
\eea
For these gauge conditions the Faddeev - Popov determinant
is a constant
\be
det\|\{\chi_\alpha,\varphi_\beta\}\| = 1
\ee
The Lagrange multipliers  $u_1$,  $u_2$, can be fixed from the
requirement of the  stationarity of a gauge condition  under the
time evolution governed by the extended  Hamiltonian
$$
H_E = \frac{1}{2}(p_1^2 + p_2^2)-y(x_1p_2-x_2p_1) + V(x_1^2+x_2^2)+
u(t)_1 p_y.
+ u_2 (x_1p_2-x_2p_1)$$
\bea
&&u_1 =0 \nn\\
&&u_2 = y - \frac{x_1p_2 - x_2p_1}{\sqrt{x_1^2 + x_2^2}}\nn
\eea
Thus the gauge (\ref{eq:cclg}) obey condition
of attainability ( intersect all gauge orbits )
and fix the gauge freedom in unique way
and leads to the dynamics equivalent to the dynamics obtained
by gaugeless method (\ref{eq:mheq}).

Now let us consider  the following gauge condition
\bea \label{eq:gch}
&&\chi_1 \,\equiv \,y = 0 \nn \\
&&\chi_2 \,\equiv \,\frac{x_1^2 -x_2^2}{x_1^2 + x_2^2} -
\left ( \frac{1}{2} +  \frac{A}{\sqrt{x_1^2 + x_2^2}} \right ) = 0,
\quad A > 0.
\eea
One can verify that for the gauge (\ref{eq:gch}) there is a an
obstruction to fulfilling
the requirement  of attainability.
Indeed in terms of special coordinates (\ref{eq:lctr})
these gauges look as
\bea \label{eq:sgch}
&&\chi_1 \equiv Y = 0 \nn \\
&&\chi_2 \equiv \cos 2\overline\Theta -
\left ( \frac{1}{2} +  \frac{A}{R} \right ) = 0,
\eea
 From (\ref{eq:sgch}) we have \( 0< 2\overline\Theta \leq \frac{\pi}{3}\, ,
 \frac{5\pi}{3} \leq 2\overline\Theta < 2\pi \)
and thus the nonsingularity of the Faddeev - Popov determinant
on the physical submanifold \( \Gamma^\ast \) is fulfilled
\be
det\|\{\chi_\alpha,\varphi_\beta\}\|= {-2\sin 2\overline\Theta}
 \Bigl\vert_{\Gamma^\ast}
\not=0
\ee
One can again fix the Lagrange multipliers  $u_1$,  $u_2$,
and get the description of dynamics of reduced system
but it will bee
non equivalent to the dynamics obtained
by gaugeless method (\ref{eq:mheq}) owing to the obvious
in special coordinates restriction on physical variable
\(R\)
\be
R > 2A
\ee
Finally one could note that the proposed condition
for admissible gauges (\ref{eq:suffcond}) forbid the using of
this type of gauges.

\subsection{\sf Non - Abelian Christ \& Lee model  }
\bigskip

\begin{center}
{\it Abelianization of constraints}
\end{center}

In this section we will apply  the above described procedure
of abelianization of constraints to
the well known example; non - Abelian  Christ and Lee model
described by  Lagrangian
\[
{\cal L} ({\bf x}, \dot{\bf x}, {\bf y}) =
\frac{1}{2}\left(\dot{\bf x}- [{\bf y},{\bf x}] \right)^2 - V({\bf x}^2)
\]
where \({\bf x}\) and \({\bf y}\) -are  the three - dimensional vectors,
(\(x_1, x_2, x_3)
,(y_1, y_2, y_3\)).

It is easy to verify that except for three primary constraints
\[
{\bfpi} = \frac{\partial{\cal L}}{\partial{\dot{\bf y}}} = 0
\]
there are  two independent  constraints
\bea
\Phi^0_1 &= & x_2 p_3 -x_3 p_2 \nn\\
\Phi^0_2 & =& x_3 p_1 -x_1 p_3
\eea
with the algebra
\bea
\{\Phi^0_1 ,\Phi^0_2 \} &= &
- \frac{x_1}{x_3}\Phi^0_1 - \frac{x_2}{x_3}\Phi^0_2
\eea
The abelianization procedure for this simple case consist of two stages
At the first step the transformation \({\cal S}^1\) reduces to the
\bea
\Phi^1_1 & = & \Phi^0_1 \nn\\
\Phi^1_2  & = & \Phi^0_2  + C\Phi^0_1
\eea
and equation (\ref{eq:lin1}) looks like
\bea
\{\Phi^0_1 ,C \} & =&
 \frac{x_2}{x_3} C + \frac{x_1}{x_3}
\eea
One can write down a particular  solution for  this equation
\be
C(x) =  \frac{x_1}{x_3} \arctan{\left( \frac{x_2}{x_3} \right)}
\ee
So, as a result of the first step we get a new algebra
\bea
\{\Phi^1_1 ,\Phi^1_2 \} &= &
- \frac{x_2}{x_3}\Phi^1_2
\eea
Now let us perform the  second transformation \( {\cal R}^2 \)
\bea
\Phi^2_1 & = & \Phi^1_1 \nn\\
\Phi^1_2  & = & B \Phi^1_2
\eea
with the function \(B \) that satisfies the equation of type
(\ref{eq:lin2})
\bea
\{\Phi^1_1 ,B \} & =& \frac{x_2}{x_3}
\eea
A particular  solution for  this equation reads
\be
B(x) = \ln \left( \frac{\sqrt{x_2^2+x_3^2}}{x_3} \right )
\ee
Thus, the Abelian constraints equivalent to the initial non - Abelian ones
have  the form
\bea
\Phi^2_1 & = & x_2 p_3 -x_3 p_2  \\
\Phi^2_2  & = & \ln \left(\frac{\sqrt{x_2^2+x_3^2}}{x_3} \right )
\left[ (x_3 p_1 -x_1 p_3) +
 \frac{x_1}{x_3} \arctan{\left( \frac{x_2}{x_3} \right)}
(x_2 p_3 -x_3 p_2)
\right]\nn
\eea

\bigskip

\section{\sf Concluding remarks}

To separate  the true dynamical variables from the nonphysical ones
in the  classical Hamiltonian systems with first class
constraints
without  any gauge  fixing, we have developed the gaugeless approach.
In this approach,  the reduced phase space is constructed without gauge
fixing condition using  the  procedure of local abelianization of
constraints  with the subsequent  canonical transformation
so that some of the new momenta which
are equal to the new abelian constraints while the corresponding conjugate
coordinates are ignorable (nonphysical) one. The remaining  canonical
pairs form the basis of the reduced phase space.
We have discussed the gauge fixing and gaugeless  methods for reducing the
phase space  of a singular system with the aim to  study  the problem of
determination of admissible gauges .
We have  introduced the notion of canonical gauges as functions
depending only on the nonphysical variables.
It is interesting to notice that suggested condition
(\ref{eq:suffcond})
for a gauge to be a canonical
has a simple geometric meaning. As it is know  \cite{Nester},
the inverse
of Faddeev - Popov matrix \(\Delta^{-1}\) represents the element of
volume of phase space
\( \overline{\Gamma} \equiv \Gamma \setminus \Gamma^\ast \) written
in noncanonical coordinates, and thus our condition
(\ref{eq:suffcond})  means its conservation
in the process of the time evolution.

The final goal of our consideration is the
construction of the reduced phase space for the complicated
non - Abelian gauge theory and gravity. This program is presently
under investigation, and the current article is the first step
in this direction.
The application of our scheme to the  SU(2) Yang - Mills  will be
done in separate forthcoming publication.

\bigskip

\section{\sf \,\,Acknowledgments}
\bigskip

We are happy to acknowledge interesting and critical
discussions with Professors B.M. Barbashov, A.Dubin, A.T. Filippov,
A.N. Kvinikhidze,
 G.Lavrelash\-vili, V.V.Nesterenko, V.A. Rubakov, A. Wipf.
One of us (A.M.K.)  would like to thank Prof. D.Wyller for kind hospitality
at the Institute  for Theoretical Physics of Zurich University
were part of this work was done.
This work was
supported in part by the Russian Foundation  of Fundamental Investigations,
Grant No 95\--02\--14411.
Work of A.M.K. was partly supported
by the Swiss National Foundation.

\end{document}